\documentclass[twocolumn,floatfix,10pt,aps,pre]{revtex4-1}
\usepackage{ulem}
\usepackage{amssymb,amsfonts,amsmath,epsfig,color}
\usepackage{hyperref}

\newcommand{\bvec}[1]{{\mathbf{\string#1} }}
\newcommand{\upd}{\mathrm{d}}
\newcommand{\fu}[2]{\ensuremath{\left\{\begin{array}{c}#1\\#2\end{array}\right.}}

\newcommand{\Vext}{V_\text{ext}}
\newcommand{\dt}{\Delta t}
\newcommand{\Ncanmix}{N_1 N_2}

\begin{document}

\title{Particle-conserving dynamics on the single-particle level}
\author{T.\ Schindler}\affiliation{Institute for Theoretical Physics I, Friedrich-Alexander University Erlangen-N\"urnberg, DE-91058 Erlangen, Germany}
\affiliation{Theoretical Physics II, University of Bayreuth, DE-95444 Bayreuth, Germany}
\author{R.\ Wittmann}
\affiliation{Department of Physics, University of Fribourg, CH-1700 Fribourg, Switzerland}
\author{J.\ M.\ Brader}\affiliation{Department of Physics, University of Fribourg, CH-1700 Fribourg, Switzerland}
\date{\today}

\begin{abstract}
We generalize the particle-conserving dynamics method of de las Heras {\it et al.}\ [J.\ Phys.\ Condens.\ Matter:\ {\bf 28}, 24404 (2016).] to binary 
mixtures and apply this to hard rods in one dimension. 
Considering the case of one species consisting of only one particle enables us to address the tagged-particle dynamics. 
The time-evolution of the species-labeled density profiles is compared to exact Brownian dynamics and (grand-canonical) dynamical density functional theory.
The particle conserving dynamics yields improved results over the dynamical density functional theory and well reproduces the simulation data at short and 
intermediate times.
However, the neglect of a strict particle order (due to the fundamental statistical assumption of ergodicity) leads to errors at long times for our one-dimensional setup.
The isolated study of that error makes clear the fundamental limitations of (adiabatic) density-based theoretical 
approaches when applied to systems of any dimension for which particle caging 
is a dominant physical mechanism.
\end{abstract}

\maketitle

\section{Introduction}
Dynamical density functional theory (DDFT) is a widely used and versatile tool for 
investigating the dynamics of bulk and inhomogeneous classical systems of interacting 
Brownian particles. 
By assuming that all pair and higher-order correlation functions equilibrate much faster than 
the one-body (or one-point) density (an adiabatic approximation) the DDFT exploits the formally exact 
statistical mechanical method of density functional theory (DFT) to approximately treat 
nonequilibrium situations~\cite{marconi_tarazona}. 

DDFT has been applied with much success to study a variety of important physical phenomena, 
e.g.~spinodal decomposition~\cite{archer_evans}, colloidal sedimentation~\cite{royall} 
and quasicrystal formation~\cite{quasi}. 
When combined with the test-particle method, whereby one of the particles is treated as 
an external field, DDFT can be used
to calculate the self and distinct parts of the van Hove function~\cite{dtp,stopper_bulk} 
and thus address the dynamics of equilibrium states. 
Extensions of the theory to treat driven systems can reproduce the phenomenology of 
colloidal system under external shear flow~\cite{bk1,scacchi,stopper_shear}. 
More recently a general and exact variational framework, power functional theory,
has been developed to treat nonequilibrium Brownian systems 
\cite{pft1}. 
Within this approach, DFT and DDFT emerge as equilibrium and adiabatic limits, 
respectively. 
Moreover, power functional theory enables the superadiabatic contributions to the dynamics to be approximated 
in a physically intuitive way~\cite{pft2}. 

An implicit drawback to standard implementations of DDFT is that interaction 
forces are generated from a grand-canonical free energy functional. 
For confined systems with small numbers of particles or, perhaps more generally, 
systems which exhibit strong density inhomogeneities, grand-canonical DFT can 
predict density profiles which differ significantly from canonical simulations 
at fixed particle number $N$. 
This issue was addressed in the late 1990's by Gonz\'alez {\it et al.}~\cite{canonical1,canonical2}, who employed an expansion in 
inverse powers of the particle number to systematically approximate canonical density profiles 
using grand-canonical information as an input. 
The problem of calculating equilibrium density profiles in the canonical ensemble 
was revisited in 2014 by de las Heras and Schmidt~\cite{canonical3} who showed 
how to obtain exact canonical information from grand-canonical DFT 
given an exact functional, performing 
explicit calculations for the one-dimensional hard-rod system. 
The method of de las Heras and Schmidt was further generalized in 
Ref.~[\onlinecite{canonical4}] to generate a theory of particle-conserving dynamics (PCD)
for the time-evolution of the one-point density of $N$ particles. 
This theory, while still relying on the adiabatic approximation, eliminates spurious 
effects arising from the grand-canonical ensemble and yields predictions for 
the time-dependent density in good agreement with Brownian dynamics (BD) simulation data. 

In this paper we provide an intuitive generalization of the PCD approach of Ref.~[\onlinecite{canonical4}] to binary
mixtures and use this to examine the dynamics of a tagged particle in a one-component 
system. 
By tagging a particle we can investigate the physics of dynamical confinement or ``caging'' 
within the framework of an adiabatic time-evolution equation for the one-point density. 
For clarity in our terminology we 
only speak here of a ``one-point density'', as opposed to two- or $n$-point densities (correlations),
instead of using the equivalent term ``one-body density'', which should not be confused with the particular (canonical)
density for $N=1$ particle, i.e., a single-particle density profile.
Our calculations are performed for hard-rods in one spatial 
dimension because (i) the exact grand-canonical density functional is known
and (ii) the Tonks gas~\cite{tonks} provides an extreme case of a non-ergodic fluid, which allows us to illustrate most clearly the physics of localization and caging. 
Finally, (iii), the dynamic state of hard rods constitutes a fundamental model for single-file diffusion of particles that never swap positions, 
which can be solved in a closed form~\cite{LA1,LA2} but
not in the context of a variational framework.

The paper is organized as follows. 
In Sec.~\ref{sec_varGEN} we describe the transformation from grand-canonical to canonical information 
for mixtures and discuss the statistical background of hard rods in one dimension.
Then we embed in Sec.~\ref{sec_pcd} the canonical free energy functional in a dynamical framework
by making an adiabatic approximation of DDFT.
Then we apply this PCD approach to the relaxation dynamics of single-particle profiles of hard rods
and discuss in detail the similarities and differences to BD results.
Our findings are discussed in Sec.~\ref{sec_discussion}.

\section{Variational calculus for canonical mixtures \label{sec_varGEN}}
The variational character of DFT is fundamental to its usefulness in 
addressing the physics of the liquid state.
However, for this to apply it is necessary 
to work in the grand-canonical ensemble. 
We thus adapt
the recent method developed by de las Heras and Schmidt to use the grand-canonical data obtained from DFT, 
specifically the one-point density profiles and partition functions, to calculate the 
one-point density profile of one-component canonical systems, for which the particle number 
does not fluctuate~\cite{canonical4}. 
Here we show that it is straightforward to generalize the grand-canonical--canonical 
transformation method to binary (or $\kappa$-component) mixtures.

\subsection{Canonical information from a grand-canonical theory \label{sec_inversion}}

Consider a two-component system containing $N_\nu$ particles of species $\nu\in\{1,2\}$.
Suppose we have total grand-canonical information on such a system, i.e.,
for the given chemical potentials $\mu_\nu$, we know the grand partition function $\Xi(\mu_{1},\mu_{2})$ and the grand-canonical one-point density profiles $\rho^{(\nu)}(x)$ of each species $\nu$ at position $x$.
Formally, these can be obtained by an infinite summation of the canonical partition function $Z_{\Ncanmix}$ and one-point density $\rho_{\Ncanmix}^{(\nu)}(x)$
according to
\begin{equation}
\Xi(\mu_{1},\mu_{2})=\sum_{N_{1}=0}^{\infty}\sum_{N_{2}=0}^{\infty}e^{\beta(\mu_{1}N_{1}+\mu_{2}N_{2})}Z_{\Ncanmix}\,.\label{eq_partition_sum}
\end{equation}
and
\begin{equation}
\rho^{(\nu)}(x)=\sum_{N_{1}=0}^{\infty}\sum_{N_{2}=0}^{\infty}p_{\Ncanmix}(\mu_{1},\mu_{2})\rho_{\Ncanmix}^{(\nu)}(x)\,,\label{eq_partition_rho}
\end{equation}
where $\beta=(k_\text{B}T)^{-1}$ denotes the inverse temperature with Boltzmann's constant $k_\text{B}$ and the probability $p_{\Ncanmix}(\mu_{1},\mu_{2})$
to find a state with $N_{1}$ and $N_{2}$ particles at a given pair of chemical potentials $(\mu_{1},\mu_{2})$ is given by 
\begin{equation}
p_{N}(\mu_{1},\mu_{2})=e^{\beta(\mu_{1}N_{1}+\mu_{2}N_{2})}\frac{Z_{\Ncanmix}}{\Xi(\mu_{1},\mu_{2})}\,.\label{eq_probabilities}
\end{equation}
In practice, the sums can be truncated at the maximum particle number $N_{\text{max}}$ in finite systems, i.e., all partition sums and probabilities with $N_{1}+N_{2}> N_{\text{max}}$ vanish.
Otherwise, there is no difference between the two ensembles in the thermodynamic limit.

We calculate the grand-canonical density
distributions and partition functions
to set up a system of 
$M\equiv(N_{\text{max}}+1)(N_{\text{max}}+2)/2$
linear equations in the form of Eq.~(\ref{eq_partition_sum}), with $M$ being the number of possible pairs of particle numbers, $(N_{1},N_{2})$, with $N_{1}+N_{2}\leq N_{\text{max}}$.
We then solve this equation system for $Z_{\Ncanmix}$ and calculate
the probabilities $p_{\Ncanmix}(\mu_{1},\mu_{2})$ via Eq.~(\ref{eq_probabilities}).
By interpreting these probabilities as entries of an $M\times M$
matrix $\mathbf{P}_{\Ncanmix}(\mu_{1},\mu_{2})$
and inverting this matrix we finally obtain the canonical densities via 
\begin{equation}
\rho_{\Ncanmix}^{(\nu)}(x)=\sum_{(\mu_{1},\mu_{2})}^{M}\mathbf{P}_{\Ncanmix}(\mu_{1},\mu_{2})^{-1}\rho^{(\nu)}(x)\,.
\label{eq_inversion}
\end{equation}
The sum on the right-hand side includes $M$ arbitrary pairs of chemical potentials $\mu_{1}$ and $\mu_{2}$ of the two components.
The explicit choice does not matter given the available grand-canonical information is exact.
For the sake of numerical robustness, a certain range of resulting average particle numbers should be covered.
As detailed in Ref.~[\onlinecite{canonical3}], we could reduce the number of equations by one upon removing the trivial case of zero particles in each species,
which has been omitted in the present study due to the negligible effect on computation time.

The above methods can be easily generalized to systems containing any number $\kappa$ of different components,
where the number of coupled linear equations to be solved grows exponentially with $\kappa$.

It is important to notice that the density profiles and partition sums depend explicitly on the numbers of particles of each species and not only on the total particle number $N=N_1+N_2$, as soon as the species are physically distinguishable (by their particle-particle interactions or by interactions with external potentials). However, we will consider systems with fixed values of $N_{1}$ and $N_{2}$ in the remainder of the paper and hence we will indicate quantities in canonical ensembles by $N$ to unclutter the notation. Quantities with two indices $N_{1}$ and $N_{2}$ will then indicate ordered ensembles (see Sec.~\ref{sec_distinguish}).

\subsection{Classical density functional theory\label{sec_DFT}}

Our starting point is classical DFT, providing, via a variational formalism, the grand partition function and the
grand-canonical equilibrium density profiles of an arbitrary mixture of particles under the influence of any external field $\Vext^{(\nu)}(x)$ acting on each component $\nu$.
Given a density functional $\Omega[\{\rho^{(\nu)}\}]$ of the grand potential, one obtains the grand-canonical equilibrium density profiles of each component from a variational minimization according to
\begin{align}
\frac{\delta\Omega[\{\rho^{(\nu')}\}]}{\delta\rho^{(\nu)}(x)}=0\,.
\end{align}
Substituting the resulting equilibrium density profile 
into the functional yields the equilibrium grand potential $\Omega$ and, therefore, 
the grand partition function $\Xi=\exp(-\beta\Omega)$.
Canonical information is then accessible via inversion of Eqs.~(\ref{eq_partition_sum}) and~(\ref{eq_partition_rho}). 

Here, we restrict ourselves to hard rods of length $\sigma$ in one spatial dimension, where the exact density functional reads
\begin{align} 
\Omega[\{\rho^{(\nu)}\}]=\mathcal{F}_\text{id}
+\mathcal{F}_\text{ex}
+\sum_{\nu=1}^2\int\upd x\,\rho^{(\nu)}(x)\left(\Vext^{(\nu)}(x)-\mu_\nu \right).
\label{eq_DFrho}
\end{align}
with the contribution 
\begin{align} 
\beta\mathcal{F}_\text{id}[\{\rho^{(\nu)}\}]=\sum_{\nu=1}^2 \int\upd x\,\rho^{(\nu)}(x)\, (\ln(\Lambda\rho^{(\nu)}(x))-1)
\label{eq_DFrhoID}
\end{align}
of an ideal non-interacting gas, where $\Lambda$ is the thermal wavelength.
The excess free energy functional of hard-rods was derived by Percus~\cite{percus,percusmix} 
and is given by
\begin{align} 
\beta\mathcal{F}_\text{ex}[\rho]
=-\int\upd x\, n_0(x)\ln\left(1-n_1(x)\right).
\label{eq_DFrhoEX}
\end{align} 
This functional describes the contribution of the pair interaction potential 
\begin{align} 
u(x)=\fu{0\ \ \ \ \, \text{if}\ |x|>\sigma}{\infty\ \ \ \text{else}\hfill}
\label{eq_u}
\end{align}
and is a function of the two weighted densities
\begin{align} 
n_i(x)&= \sum_{\nu=1}^2 \int\upd x'\,\rho^{(\nu)}(x')\,\omega^{(i)}(x-x')\,,
\label{eq_DFrhoEWD}
\end{align}
where $i$ is not a species label but serves to enumerate the weighted densities.
The corresponding weight functions are given by 
\begin{align} 
\omega^{(0)}(x)&=\frac{1}{2}\left( \delta(R-x) + \delta(R+x) \right)
\\ 
\omega^{(1)}(x)&=\Theta(R-|x|),
\end{align}
where $R=\sigma/2$ is half the length of a rod
and $\delta(x)$ and $\Theta(x)$ denote the Dirac distribution and the Heaviside step function, respectively. 
Here we dropped all species labels, since we only consider identical particles (in general, the functional allows for the description of $\kappa$ actually different species of lengths $2R_\nu$).
For the virtual mixture considered here, we further use the notation $\mathcal{F}_\text{ex}[\rho]$
to emphasize that Eq.~\eqref{eq_DFrhoEX} then depends only on the total density profile $\rho(x)=\rho^{(1)}(x)+\rho^{(2)}(x)$.

\subsection{Canonical intrinsic free energy functional \label{sec_canonicalDFT}}

Although there exists an explicit expression for a density functional $\mathcal{F}^\text{tot}=\Omega+\sum_\nu\int\upd x\,\rho^{(\nu)}(x)\,\mu_\nu$ of the total Helmholtz free energy, cf.\ Eq.~\eqref{eq_DFrho},
its minimization under the constraint of fixed particle numbers $N_\nu=\int\upd x \rho^{(\nu)}(x)$ of each component
would still result in a grand-canonical profile, i.e.,
a superposition of canonical density profiles with different particle numbers that average to $N_\nu$.
This constraint is exactly what requires one to introduce chemical potentials.
Our objective is rather to provide a true canonical DFT of the form
\begin{align}
F^\text{tot}_{N}[\{{\rho_N^{(\nu)}}\}]= F_{N}[\{{\rho_N^{(\nu)}}\}]+\sum_{\nu=1}^2\int\upd x\,{\rho_N^{(\nu)}}(x) \Vext^{(\nu)}(x)
\label{eq_FintrTOT}
\end{align}
where the total Helmholtz free energy is formally written in its natural variables, which implies that all valid canonical ``target'' profiles ${\rho_N^{(\nu)}}(x)$ must integrate to $N_\nu$
and the functional must be minimal in canonical equilibrium.
To this end we must perform an iterational search for the intrinsic Helmholtz free energy functional $F_{N}$ on the right-hand side of Eq.~\eqref{eq_FintrTOT}.

In generalization of one-component case described in Ref.~[\onlinecite{canonical4}] we determine for a given pair ${\rho_N^{(\nu)}}(x)$ of ``target'' profiles 
the corresponding generating external potentials $V^{(1)}(x)$ and $V^{(2)}(x)$ acting on each species such that the target profiles would be equilibrated.
Then the intrinsic Helmholtz free energy functional is obtained as
\begin{align}
\beta F_{N}[\{{\rho_N^{(\nu)}}\}]=-\ln Z_{N}-\sum_{\nu=1}^2\int\upd x\,{\rho_N^{(\nu)}}(x)\beta V^{(\nu)}(x)\,.
\label{eq_FintrCAN}
\end{align}
To determine $V^{(\nu)}(x)$ for a given canonical target profile, we start with an initial guess 
for $V_0^{(\nu)}(x)$ and then employ the gradient-free iteration scheme~\cite{canonical4}
\begin{align} 
\beta V^{(\nu)}_{n}(x)=\beta V^{(\nu)}_{n-1}(x)-\ln{\rho_N^{(\nu)}}(x)+\ln\rho^{(\nu)}_{n-1}(x)\,.
\label{eq_itV}
\end{align}
In each iteration step $n$ we make use of the canonical equilibrium profile $\rho_{n-1}^{(\nu)}(x)$ of species $\nu$ in the external potential $V^{(\nu)}_{n-1}(x)$ of the previous step.
It is found by minimizing Eq.~\eqref{eq_DFrho} with $V^{(\nu)}_{n-1}(x)$ taking the role of $\Vext^{(\nu)}(x)$ and then
inverting Eq.~\eqref{eq_partition_rho} as described in Sec.~\ref{sec_inversion}.
In practice, we ensure a proper convergence of Eq.~\eqref{eq_itV} by introducing a damping factor close to unity, lowering the weight of the logarithmic terms, 
and adding a very small number to the density, avoiding a divergence of the logarithm in case of vanishing density.
Note that the canonical partition function changes in each iteration step and generally differs from $Z_{N}$ entering in Eq.~\eqref{eq_FintrCAN}, which is calculated for the true external fields $\Vext^{(\nu)}(x)$
and that the obtained $V^{(\nu)}(x)$ are unique only up to an irrelevant constant related to the initial guess.

In the convention chosen here, the potentials $V^{(\nu)}(x)$ formally replace $\Vext^{(\nu)}(x)$ but still take into account the constraints given by these actual external fields, since, otherwise, Eq.~\eqref{eq_FintrTOT} would be ill-defined.
Hence, we can separate $V^{(\nu)}(x)$ into $\Vext^{(\nu)}(x)$ and a ``nonequilibrium'' correction, i.e., due to considering nonequilibrium target profiles ${\rho_N^{(\nu)}}(x)$.
To illustrate this, we demonstrate how to minimize the canonical free energy functional, Eq.~\eqref{eq_FintrTOT}, by setting
$\delta F^\text{tot}_{N}/\delta {\rho_N^{(\nu)}}(x)=0$.
The resulting condition
\begin{align}
\Vext^{(\nu)}(x)-V^{(\nu)}(x)-\sum_{\nu'=1}^2\int\upd x'\,{\rho_N^{(\nu)}}(x')\frac{ \delta V^{(\nu')}(x')}{\delta {\rho_N^{(\nu)}}(x)}=0\,
\end{align}
is satisfied by $\Vext^{(\nu)}(x)=V^{(\nu)}(x)$,
which can be formally solved for $\rho_{N}^{(\nu)}(x)$ by an iteration similar to Eq.~\eqref{eq_itV}.
In practice, such a calculation would just amount to determine the canonical equilibrium profile in the external field $\Vext^{(\nu)}(x)$ indirectly
by one single grand-canonical-canonical transformation according to Eq.~\eqref{eq_inversion}.

\begin{figure}
\begin{centering}
\includegraphics[width=0.4\textwidth]{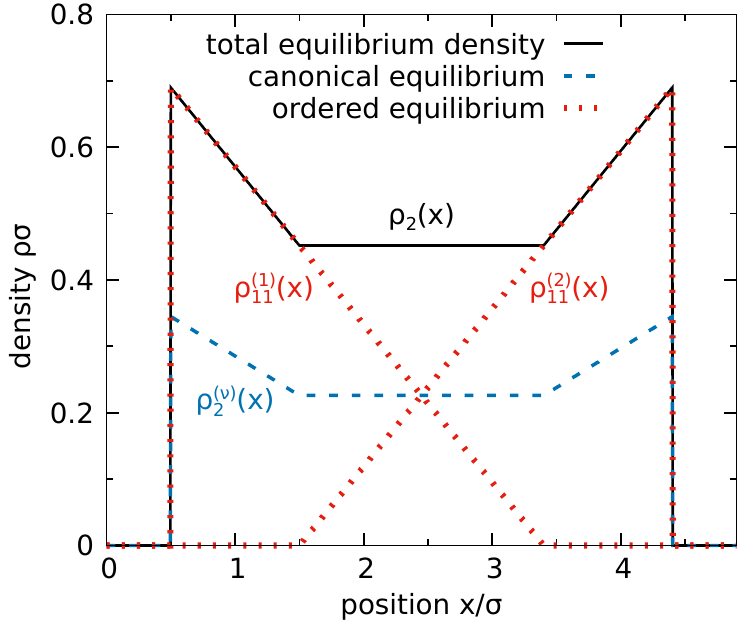}
\par\end{centering}
\protect\caption{(color online)
Equilibrium density profiles of two hard rods of length $\sigma=2R$ in a one dimensional slit of length $L=4.9\sigma$. 
Solid line: the total canonical density profile $\rho_2(x)$ for one component, cf.\ Eq.~\eqref{eq_rho11eqNO}.
Dotted lines: species-resolved density profiles for a mixture in which the particle order is strictly maintained, 
as given by Eq.~\eqref{eq_rho11eq}.
Dashed line: species-resolved canonical densities as given by Eq.~\eqref{eq_rho11eqNO};
since the canonical ensemble does not respect particle ordering the two curves are identical. 
\label{fig_N2eq}}
\end{figure}

\subsection{Distinguishability in one dimension \label{sec_distinguish}}

Given that the canonical transformation procedure has been demonstrated to be formally 
correct~\cite{canonical4} it is interesting to proceed to investigate the limitations 
of the canonical ensemble for describing systems subject to non-ergodic dynamics. 
Due to the constraint that the particles remain ordered on the line, the one-dimensional 
hard rod model presents one of the simplest non-ergodic model systems. 
We will first highlight the inability of the canonical ensemble to correctly describe 
species-labeled density profiles in equilibrium, before proceeding in Sec.~\ref{sec_pcd}
to consider the PCD of tagged particles. 
We will then argue that our findings have strong implications for the ability of any 
approach based on ensemble-averaged density (adiabatic or superadiabatic)
to describe non-ergodic behavior arising from particle 
caging. 

If we employ species labeling simply as a formal device to track either individual particles 
or subsets of particles, then, within a canonical description, the 
equilibrium density profiles of a species holding $N_\nu$ particles are always given by 
$\rho_{N}^{(\nu)}=N_\nu\rho_N(x)/N$, where $\rho_N(x)$ is the total canonical equilibrium density profile of 
$N$ identical particles, irrespective of their species labeling. 
This is in contradiction with the real situation in systems with broken ergodicity, such as
densely packed crystals, glasses or, in the example we lay out in the following, hard rods in 
one-dimension, where the single-particle profiles should reflect the spatial localization.

To illustrate the origin of localization in a statistical description, we analyze the simplest 
case with a non-trivial pair interaction:
$N=2$ hard rods of length $2R$
confined to a slit of length $L$. 
The canonical partition function, $Z_{2}=(L-4R)^2/(2\Lambda^2)$, of two particles
can be calculated in two different ways. 
We stress that by the word canonical we always imply the ergodic assumption, i.e., the statistical average implies no particular particle order.
The standard approach for completely indistinguishable particles is to calculate
\begin{align}
Z_{2}=\frac{1}{2!\Lambda^2}\int_{R}^{L-R}\mathrm{d}x_1\int_{R}^{L-R}
\mathrm{d}x_2\,e^{-\beta u(|x_2-x_1|)}
\label{eq_Z2}
\end{align}
via the completely symmetric pair interaction potential $u(x)$, specified in Eq.~\eqref{eq_u}.
Alternatively, if we distinguish between the two particles and require that particle 1 is always to the left of 
particle 2, then the ordered partition function reads
\begin{align}
Z_{11}\equiv\frac{1}{\Lambda^2}\int_R^{L-3R}\mathrm{d}x_1\int_{x_1+2R}^{L-R}\mathrm{d}x_2\equiv\mathcal{Z}_{11}[1,2]\,,
\label{eq_Z11}
\end{align}
thus implying a broken ergodicity.
In the last step we have introduced the formal arguments 1 and 2 to express the explicit dependence on the order of
the two particles (the alternative functional notation $\mathcal{Z}_{11}[1,2]$ implies that particle 1 lies to the left of particle 2, which is not decisive for the mathematical value of $Z_{11}$ but serves to indicate the explicit particle order).
Equivalently we can calculate
\begin{align}
Z_{11}
=\frac{1}{\Lambda^2}\int_{R}^{L-R}\mathrm{d}x_1\int_{R}^{L-R}
\mathrm{d}x_2\,e^{-\beta o(x_2-x_1)}\,,
\label{eq_Z11or}
\end{align}
where the ordering pair potential
\begin{align} 
o(x)=\fu{0\ \ \ \ \, \text{if}\ x>\sigma}{\infty\ \ \ \text{else}\hfill}\,
\label{eq_o}
\end{align}
depends on the relative distance and not on its absolute value.

It is straightforward to see that both partition functions are mathematically equal (they evaluate to the same number) and we can formally write
\begin{align}
Z_{2}
=\frac{1}{2!}\left(\mathcal{Z}_{11}[1,2]+\mathcal{Z}_{11}[2,1]\right)\,,
\label{eq_Z2Z11comp}
\end{align}
which is obviously equal to $Z_{11}$.
The point we wish to make here is that, although the value of both partition functions 
$Z_{11}$ and $Z_{2}$ is the same, the corresponding one-point densities of labeled particles, 
i.e., the ensemble average of the density operators $\hat{\rho}_\nu=\delta(x-x_\nu)$ with respect to the different probability distributions implied in Eq.~\eqref{eq_Z2} and Eq.~\eqref{eq_Z11or}, respectively, are different.

The proper calculation for mixtures via the ordered probability distribution, 
associated with $Z_{11}$, yields the 
density profiles 
\begin{align}
\rho^{(1)}_{11}(x)&=\frac{2(L-3R-x)\,\Theta(x-R)\Theta(L-3R-x)}{(L-4R)^2}\,,\nonumber\\
\rho^{(2)}_{11}(x)&=\frac{2(x-3R)\,\Theta(x-3R)\Theta(L-R-x)}{(L-4R)^2}\,,
\label{eq_rho11eq}
\end{align}
shown as the dotted lines in Fig.~\ref{fig_N2eq}.
The (expected) difference between the two profiles arises from the physical distinction due to the imposed particle order, also present in BD simulations.
In contrast, from the canonical distribution, associated with $Z_{2}$, we obtain identical results for each profile
\begin{equation}\!
\rho^{(1)}_{2}(x)=\rho^{(2)}_{2}(x)=\frac{1}{2}\left(\rho^{(1)}_{11}(x)+\rho^{(2)}_{11}(x)\right)
\equiv\frac{1}{2}\,\rho_2(x)\,,\!\!\!\!\!\!
\label{eq_rho11eqNO}
\end{equation}
shown as dashed lines in Fig.~\ref{fig_N2eq}.
In a manner of speaking, we can say that the two species-labeled profiles in Eq.~\eqref{eq_rho11eqNO} follow from an additional average of those in Eq.~\eqref{eq_rho11eq}, 
accompanied by a loss of microscopic information, whereas their sum
equals the exact total canonical profile $\rho_2(x)$ for two particles in both cases.

Note that the ordered density profiles $\rho^{(\nu)}_{11}(x)$ need to be understood as correlated averaged quantities,
so that the overlap of the profiles does not indicate an explicit crossing of their trajectories .
The decreasing probability to find a particle in a specific region rather reflects the likely presence of the other particle according to the predefined order.
There are still exclusion strict regions of exactly one particle length on one side for each profile, in addition to the confining wall,
This correlation is lost for $\rho^{(\nu)}_{2}(x)$, so that each particle could be found at a position, where the other would not fit in any more. 
In general, to detect unphysical mixing, we can use the criterion of a nonzero probability for the center of a particle penetrating the minimal region
which should remain available for the other particles to respect the predefined order.

The situation described above remains qualitatively the same if the rods are physically distinguishable, e.g., by their lengths.
The only difference is the value of the partition functions, since the factor $1/2!$ in $Z_2$ has to be removed,
so that we find for the only virtual mixture of physically indistinguishable particles $Z_{2}=\left(\mathcal{Z}_{11}[1,2]+\mathcal{Z}_{11}[2,1]\right)$, which equals $2 Z_{11}$ in contrast to Eq.~\eqref{eq_Z2Z11comp}.
This factor does, however, not affect the density profiles generalizing Eqs.~\eqref{eq_rho11eq} and~\eqref{eq_rho11eqNO} to true mixtures
and, therefore, all conclusions drawn and illustrated in the following equally apply to this more general case.

\section{Particle-conserving dynamics \label{sec_pcd}} 

In this section we use the canonical intrinsic free energy functional, Eq.~\eqref{eq_FintrCAN} to drive the dynamics of a mixture of $N=N_1+N_2$ hard rods in one dimension.
In contrast to the standard (grand-canonical) DDFT, this PCD approach operates at fixed numbers of particles of each species,
thus providing a more realistic representation of the BD of a system, which, by construction, resolves the positions of all particles at each time 
$t$ (measured in units of the Boltzmann time $t_\text{B}=\sigma^2/D_0$, where $D_0$ is the common diffusion coefficient).
Our simulation setup and the averaging process to obtain the species-resolved density profiles $\rho^{(\nu)}_{N_1N_2}(x,t)$ from BD are described in the Appendix.
By considering identical particles and choosing $N_1=1$, we can further resolve within PCD the time evolution of the probability density $\rho^{(1)}_{N}(x,t)$ associated with the (average) location of a single particle.

\subsection{Adiabatic approximation}

The crucial approximation, which allows one to employ any sort of equilibrium DFT in a nonequilibrium framework is to assume
that the correlations, i.e., the $n$-particle densities for $n>1$, at each instant of time
follow from the time-dependent one-point density in the same way as in equilibrium.
This relation is provided by an equilibrium density functional and represents an adiabatic approximation,
since we approximate the dynamics as a sequence of equilibrium states.

{In DDFT, the one-point density of each species evolves in time according to~\cite{marconi_tarazona,archer_evans}
\begin{align}
\frac{\partial \rho^{(\nu)}(x,t)}{\partial t}= D_0\frac{\partial}{\partial x}&\left(\frac{\partial\rho^{(\nu)}(x,t)}{\partial x}-\beta\bvec{f}^{(\nu)}_\text{ad}(x,t)\right.\cr&\ \ \ \ \left.+\rho^{(\nu)}(x,t)\,\frac{\partial \beta\Vext^{(\nu)}(x)}{\partial x}
\right),\ \mbox{}
\label{eq_DDFT}
\end{align}
where we do not consider an explicit driving by time-dependent external fields or non-conservative forces.
The nonequilibrium interaction force 
\begin{align}
\bvec{f}^{(\nu)}_\text{ad}(x,t)=-\rho^{(\nu)}(x,t)\frac{\partial}{\partial x}\frac{\delta\mathcal{F}_\text{ex}[\{\rho^{(\nu')}\}]}{\delta\rho^{(\nu)}(x,t)}
\label{eq_fad}
\end{align}
is approximately related to the excess free energy functional from Eq.~\eqref{eq_DFrhoEX}.
With the above choices, the whole expression in the brackets on the right-hand side of Eq.~\eqref{eq_DDFT} 
could be represented, as in Eq.~\eqref{eq_fad}, by a functional derivative of the full grand-canonical free energy functional, i.e., a local chemical potential,
thereby representing entropic, internal and external forces.
We thus describe the adiabatic time evolution of the grand-canonical one-point density.
To fix the average particle numbers of each component, we calculate the initial density profiles
such that $N_\nu=\int\upd x \rho^{(\nu)}(x)$
by accordingly choosing the chemical potentials.

Within the PCD framework, we calculate the adiabatic internal force density on species $\nu$ according to~\cite{canonical4}
\begin{align}
\beta\bvec{f}^{(\nu)}_{N,\text{ad}}(x,t)=\rho_{N}^{(\nu)}(x,t)\frac{\partial}{\partial x}\left(\beta V^{(\nu)}_\text{ad}(x,t)+\ln\rho_{N}^{(\nu)}(x,t)\right)\,.
\end{align}
These expressions follow intuitively from the (adiabatic) balance with external and entropic forces on each species or,
more formally and similar to Eq.~\eqref{eq_fad}, from the functional derivative of the canonical excess free energy $\beta F_{N}-\sum_\nu\int\upd x\rho_{N}^{(\nu)}(\ln\Lambda\rho_{N}^{(\nu)}-1)$
with $F_{N}$ given by Eq.~\eqref{eq_FintrCAN}.
In any case, it is necessary to
determine at each time step the adiabatic potentials $V^{(\nu)}_\text{ad}(x,t)$ that would generate the
instantaneous canonical density profiles $\rho_{N}^{(\nu)}(x,t)$ in equilibrium, as described in Sec.~\ref{sec_canonicalDFT}.
Having made the adiabatic approximation for the canonical system, 
we obtain from the analog to Eq.~\eqref{eq_DDFT} the time evolution equations
\begin{align}
\frac{\partial \rho_{N}^{(\nu)}(x,t)}{\partial t}=\beta D_0\frac{\partial}{\partial x}\left(\rho_{N}^{(\nu)}(x,t)\,\frac{\partial\left(\Vext^{(\nu)}(x)-V^{(\nu)}_\text{ad}(x,t)\right)}{\partial x} \right)
\label{eq_DDFTgen}
\end{align}
for a mixture in PCD.
Here it becomes clear that
we can interpret the driving forces of the dynamics as the counterforces to (the nonequilibrium part of) the forces arising from $V^{(\nu)}_\text{ad}(x,t)$,
providing a physical meaning to the construction of the adiabatic potentials.

When compared to the exact BD, the DDFT approach has, in general, three drawbacks.
(i) DDFT conserves only the average number of particles of each species. This has been corrected by our modified PCD approach.
Combining the adiabatic canonical profiles according to Eq.~\eqref{eq_partition_rho} with $p_N$ independent of time, we could also provide a PCD for a grand-canonical system
(different from DDFT~\cite{canonical4}), which will not be considered here.
Moreover, (ii) superadiabatic forces are always neglected~\cite{FortiniPRL}, and 
(iii) an inexact canonical functional, obtained through inexact grand-canonical information, results in further deviations which are difficult to quantify.
In the latter case, the time evolution will depend on the initial guess $V_0^{(\nu)}(x,t)$ in the iteration procedure of Eq.~\eqref{eq_itV}, since
the vertical offset will formally renormalize the chemical potentials employed in each iteration step.
}While it is not necessary in the present study using the exact Percus functional,
it might be possible to eliminate this problem by properly shifting the generating potentials in each iteration step or, equivalently, adapting the set of chemical potentials.
In any case, it is convenient to start with the equilibrium result $V_0^{(\nu)}(x,t)=V_\text{ad}^{(\nu)}(x,t-\Delta t)$ of the previous time step, which we also do here.
In other words, an exact functional, results in exact adiabatic PCD, as prescribed by the Percus functional of the total density profile of hard rods~\cite{canonical4}.

In the following we use the generalization to mixtures of the Percus functional,
which is exact in the grand-canonical sense, i.e., when the number of particles fluctuates and their order does not matter,
but, as discussed in Sec.~\ref{sec_distinguish}, provides incorrect information on single localized particles.
Hence, we expect that PCD for one-dimensional mixtures suffers from both (i) the adiabatic approximation and
(ii) the lack of any exact density functional for a system with strict particle order (besides in the trivial case of an ideal gas, where particle trajectories may cross also in BD).
To demonstrate the still significant advantages of PCD compared to ordinary DDFT when compared to BD,
we discuss in the following the initial relaxation dynamics of hard rods in a slit.

\begin{figure}
\begin{centering}
\includegraphics[width=0.4\textwidth]{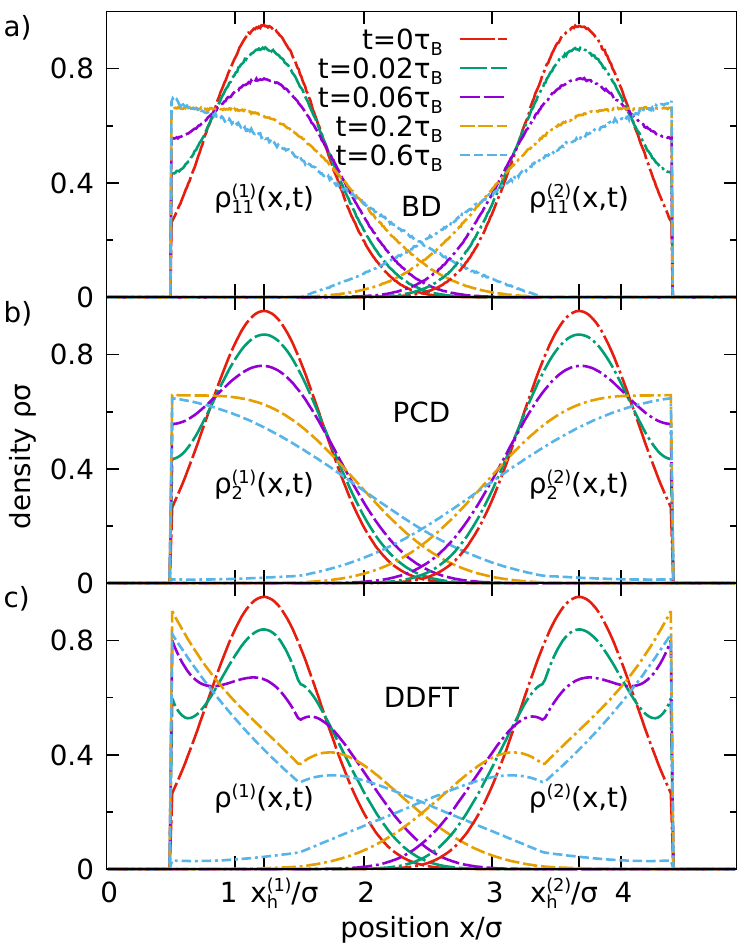}
\par\end{centering}

\protect\caption{(color online) Time evolutions of density profiles of $N=2$ hard rods in a one dimensional slit
at times indicated
for each system. We consider a two-component mixture with $N_{1}=N_{2}=1$ (dashed lines for species $\nu=1$ and dashed-dotted lines with same dash length for species $\nu=2$) and compare the results for
(a) $\rho^{(\nu)}_{11}(x,t)$ of BD simulations,
(b) $\rho^{(\nu)}_{2}(x,t)$ of our PCD approach, and (c) $\rho^{(\nu)}(x,t)$ of grand-canonical
DDFT with fixed average particle numbers of each species. 
The initial states in each case are canonical equilibrium states with an external
potential consisting of a harmonic trap $V_{\text{ext},0}^{(\nu)}(x)=k^{(\nu)}(x-x_{\mathrm{h}}^{(\nu)})^{2}/2$
for each component $\nu$ (in addition to the confining hard walls). Both harmonic traps have
a force constant of $k^{(1)}=k^{(2)}=5/\beta\sigma^{2}$ and the minima
of the external potentials are at $x_{\mathrm{h}}^{(1)}=L/4$ and
$x_{\mathrm{h}}^{(2)}=3L/4$. 
[Note, that this implies, that in (c) the same initial density profile as in (b) is used.]
At $t=0$ the traps are switched of and the density relaxes.
\label{fig_rho_of_t}}
\end{figure}

\subsection{Relaxation of the one-particle density
\label{sec_results}}

\begin{figure}
\begin{centering}
\includegraphics[width=0.4\textwidth]{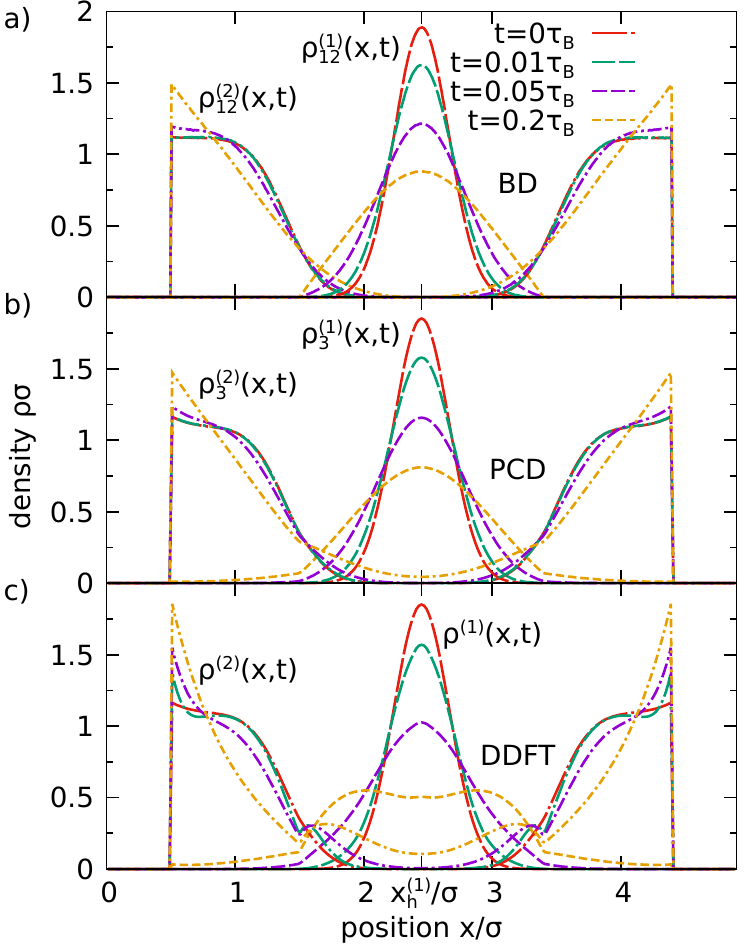}
\par\end{centering}

\protect\caption{(color online) As described in the caption of Fig.~\ref{fig_rho_of_t} but for $N_{1}=1$ and $N_{2}=2$ particles.
For the initial state only component 1 is
confined by a harmonic trap with $k^{(1)}=20/\beta\sigma^{2}$ 
and $x_{\mathrm{h}}^{(1)}=L/2$ (compare the caption in Fig.~\ref{fig_rho_of_t}), whereas there is no harmonic trap
for component 2. 
\label{fig_rho_of_tb}}
\end{figure}

To illustrate the performance of the PCD approach for a two-component mixture with a tagged particle, $N_1=1$,
we study in Figs.~\ref{fig_rho_of_t} and \ref{fig_rho_of_tb} the relaxation of the species-resolved density after bringing the system out of equilibrium by switching off
confining external potentials at $t=0$, in which the particles initially were equilibrated with respect to the chosen approach.
This initial condition ensures a high degree of localization in the density profiles,
which means that, the configurations with the wrong particle order are suppressed in the canonical DFT underlying the PCD.
To be more specific, such unphysical states are very unlikely, but not completely eliminated, 
so that there is already some small, albeit barely noticeable, difference at $t=0$ between PCD and BD with the same fixed particle numbers. 
We also compare the time evolution to DDFT, which provides grand-canonical states.
Even though not physically meaningful we start from the same initial density profiles as for the PCD calculations for better comparison of the dynamics.

In DDFT the profile is interpreted as a grand-canonical one implying repulsion within each species.
Thus the time evolution exhibits clear differences even at short times and 
it becomes apparent that PCD improves significantly over DDFT when comparing to the reference BD.

To learn more about the differences between PCD and BD,
we must look a little closer.
The PCD and BD profiles for two particles, i.e., $N_{1}=N_{2}=1$, in Fig.~\ref{fig_rho_of_t}
are quantitatively nearly indistinguishable for small times.
Hence, the neglected superadiabatic forces are insignificant in this case.
The increasing deviations emerging at $t\gtrsim0.6$ are thus of different origin.
As the particles can not exchange positions in
the one-dimensional trajectory-based BD simulation, the ordering property with particle 1 on the left-hand
side of particle 2 is conserved throughout the time evolution.
However, we clearly observe that the PCD does not conserve the order of the particles.
While the unphysical states with particle 1 on the right-hand side of particle 2 are initially suppressed,
the overlap of the density profiles of the two components grows when the system evolves in time.
This confirms our expectations based on the model calculations in Sec.~\ref{sec_distinguish}
that the present version of DFT is not able to properly drive order-preserving dynamics, and not only PCD.
In particular, the PCD profiles will ultimately approach a mixed state
without distinction between the components, i.e., the canonical equilibrium profiles depicted in Fig.~\ref{fig_N2eq}.

We also consider in Fig.~\ref{fig_rho_of_tb} the dynamics for a system with $N_{1}=1$ and $N_{2}=2$ particles with the single particle located in the middle of the slit using the same methods.
Here, for $N=3$, the role of the superadiabatic forces is a little more obvious than for $N=2$.
Regarding the time evolution of species 1, we observe best that the PCD is a little faster than BD,
which has also been observed for the one component case~\cite{canonical4} and can be generally expected from numerical simulations~\cite{FortiniPRL}.
At early times, the PCD profiles again show good agreement with the BD results.
Moreover, it does not take as long as for two particles until the the onset of the unphysical mixing occurs at $t\gtrsim0.2$.

\begin{figure}
\begin{centering}
\includegraphics[width=0.4\textwidth]{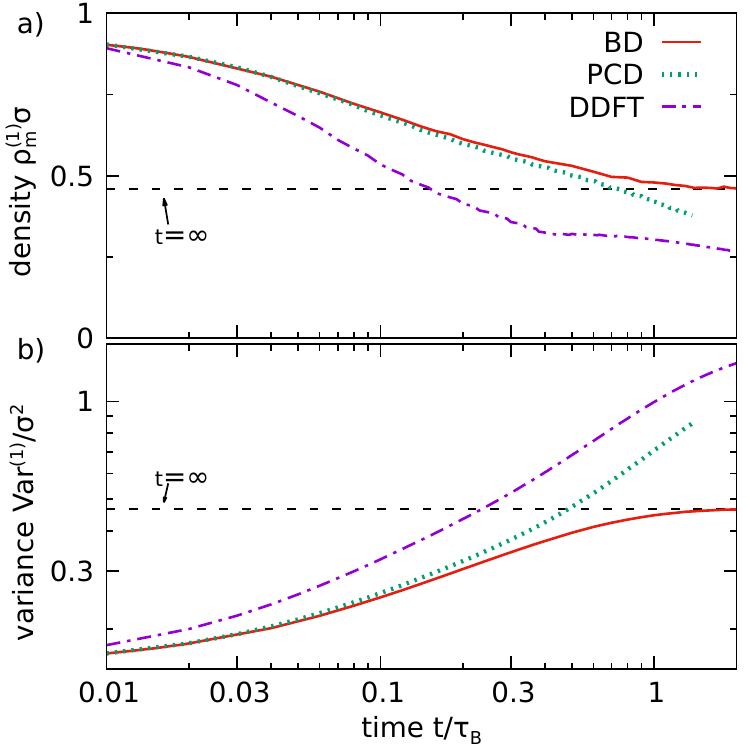}
\par\end{centering}
\protect\caption{(color online) Time evolution of mean values, as defined in the text, for the first component of the two-particle system shown in Fig.~\ref{fig_rho_of_t}.
The different methods are indicated in the legend. 
The black dashed line shows the theoretically calculated values of the
completely relaxed system in BD (at $t=\infty$).
(a) Density $\rho_{\mathrm{m}}^{(1)}$ evaluated at the mean $x$-value .
(b) Variance $\text{Var}^{(1)}(t)$ of the density distribution on logarithmic
scale. \label{fig_rhomax-variance}
}
\end{figure}

To quantify our findings we introduce a general average with respect to the one-point species-resolved densities
$\rho^{(\nu)}_{11}(x,t)$ for BD simulations, $\rho^{(\nu)}_{2}(x,t)$ for PCD and $\rho^{(\nu)}(x,t)$ for DDFT,
generally defined by 
\begin{equation}
\left\langle f(x)\right\rangle_{\nu}\equiv\frac{1}{N_{\nu}}\int\mathrm{d}x\,\rho^{(\nu)}(x)f(x)
\end{equation}
for any test function $f(x)$. 
In particular, we consider the values of the density 
$\rho_{\mathrm{m}}^{(\nu)}(t)\equiv\rho^{(\nu)}(\left\langle x\right\rangle_{\nu},t)$
evaluated at the mean $x$-coordinate $\left\langle x\right\rangle_{\nu}$
and the variance
$\text{Var}^{(\nu)}(t)\equiv\left\langle (x-\left\langle x\right\rangle_{\nu})^{2}\right\rangle_{\nu}$
as functions of time.

\begin{figure}
\begin{centering}
\includegraphics[width=0.4\textwidth]{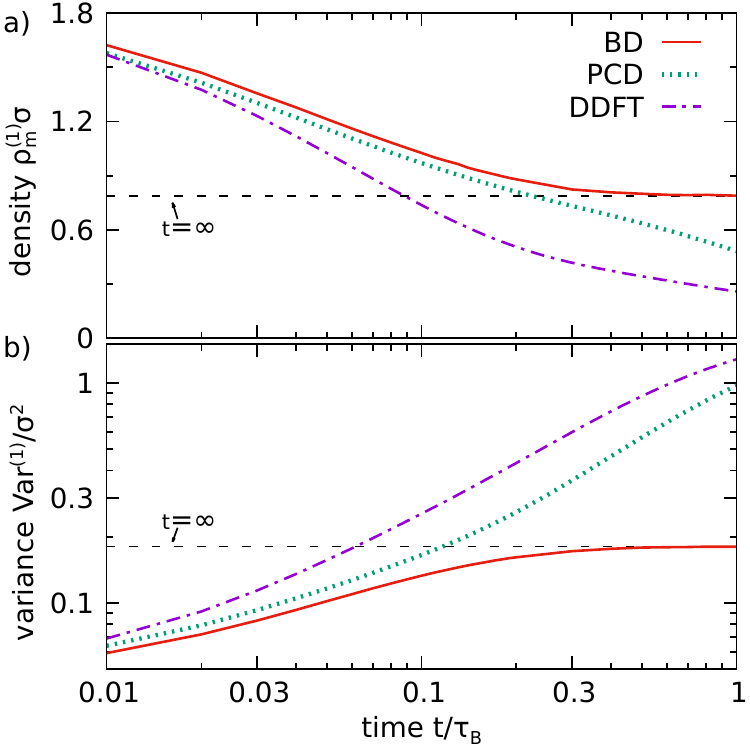}
\par\end{centering}
\protect\caption{(color online) As described in the caption of Fig.~\ref{fig_rhomax-variance},
but for component 1 of the three-particle
system shown in Fig.~\ref{fig_rho_of_tb}.
\label{fig_rhomax-varianceb}}
\end{figure}

The corresponding results $\rho_{\mathrm{m}}^{(1)}$ and $\text{Var}^{(1)}(t)$ for species $\nu=1$ are shown in Figs.~\ref{fig_rhomax-variance} and~\ref{fig_rhomax-varianceb}
for the cases $N=2$ and $N=3$, respectively, as discussed before.
In general, all observations on the different dynamics are qualitatively the same for the two systems.
The values of DDFT differ significantly from the BD simulation and PCD,
which shows that differences between canonical and grand-canonical
ensembles are important in such a small system and that the PCD provides a reliable description of the early relaxation dynamics.
Here, we also observe the one apparent difference between Figs.~\ref{fig_rhomax-variance} and~\ref{fig_rhomax-varianceb}.
For $N=3$, the PCD has both a clearly lower peak density and a greater variance than BD already at $t=0$. 
This reflects the importance of choosing some convenient
initial conditions, which are less restrictive here than for $N=2$, since the two particles of
species $\nu=2$ are not confined in a trap (cf.\ the caption of Fig.~\ref{fig_rho_of_tb})
so that they can more easily interchange with the particle in between.
This is also the reason why the unphysical mixing becomes apparent at earlier times in Fig.~\ref{fig_rho_of_t} than in Fig.~\ref{fig_rho_of_tb}.

For larger times, the studied average quantities in the BD simulations eventually reach a plateau,
whereas both PCD and DDFT continue to evolve, although the BD are initially slower due to superadiabatic forces.
In particular, the value of $\rho_{\mathrm{m}}^{(1)}(t)$ decreases further while $\text{Var}^{(1)}(t)$ continues to grow.
This illustrates that the density distributions become wider
and flatter than in BD, which is consistent with the mixing of particles.
It is clear that the overall timescale for spreading over the whole available system is larger
than for the more localized particles in BD.

\subsection{An \textit{inverse} mixing paradox for the PCD of two hard rods \label{sec_iG}}

To better understand the underlying mechanism which leads to the unphysical mixing in the PCD 
and to emphasize that it does not solely occur as a consequence of having chosen improper, already slightly mixed, initial conditions deviating from those of BD, we consider now a more extreme example.
We initiate the system in the true equilibrium state $\rho_{11}^{(\nu)}(x)$, as given by Eq.~\eqref{eq_rho11eq} and shown in Fig.~\ref{fig_N2eq}, respecting the order of two identical hard rods in a slit.
Of course, in the BD case, the density profiles will not change over time, since they are already equilibrated.

For the PCD, let us first consider the one-component case, where the (grand-canonical) 
density functional depends only on the total density $\rho(x)=\rho^{(1)}(x)+\rho^{(2)}(x)$ 
of two species and different components can be introduced on a formal level. 
In this case, the ideal free energy functional from Eq.~\eqref{eq_DFrhoID} has to be replaced with
\begin{align} 
\beta\mathcal{F}_\text{id}[\rho]=\int\upd x\,\rho(x)\, (\ln[\Lambda\rho(x)]-1)\,,
\label{eq_DFrhoIDonecomp}
\end{align}
whereas the excess free energy, Eq.~\eqref{eq_DFrhoEX} depends on the total density only, even for a mixture of identical particles.
Using this functional as the basis for the PCD, it does not matter, at any time, which particle we associate with species 1 or 2.
So, formally speaking, each pair of grand-canonical (canonical) density profiles for two species (particles), which sum up to the total density are valid adiabatic or equilibrium distributions.
In particular, there is no difference between imposing either of the two pairs of one-particle profiles in Eq.~\eqref{eq_rho11eq} or Eq.~\eqref{eq_rho11eqNO},
corresponding to the true canonical equilibrium result of two distinguishable particles or
the equilibrium result of the canonical DFT from Sec.~\ref{sec_canonicalDFT}, respectively.
In both cases, the system remains in equilibrium and the PCD approach is correct.

\begin{figure}
\begin{centering}
\includegraphics[width=0.4\textwidth]{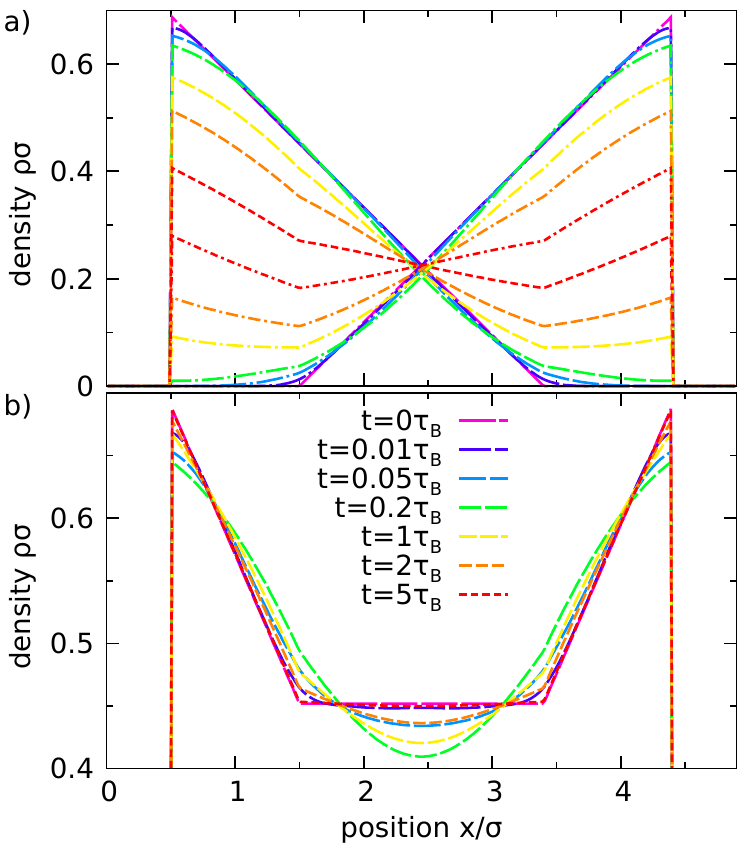}
\par\end{centering}
\protect\caption{(color online) PCD of a mixture of $N=2$ particles initialized in the ordered equilibrium state, given by Eq.~\eqref{eq_rho11eq} (the corresponding BD are time independent). 
(a) Time evolution of the single-particle profiles towards the canonical equilibrium state given by Eq.~\eqref{eq_rho11eqNO}. Dashed lines are component 1, dashed-dotted lines are component 2 with same colors (dash lengths) at the same times.
(b) The total density profile of both particles, calculated from adding up the data in (a), first moves out of and then reenters the canonical equilibrium state. 
The PCD of a one-component fluid with two particles~\cite{canonical4} would remain in the equilibrium state at $t=0$.
\label{fig_N2dyn}}
\end{figure}

\begin{figure}
\begin{centering}
\includegraphics[width=0.4\textwidth]{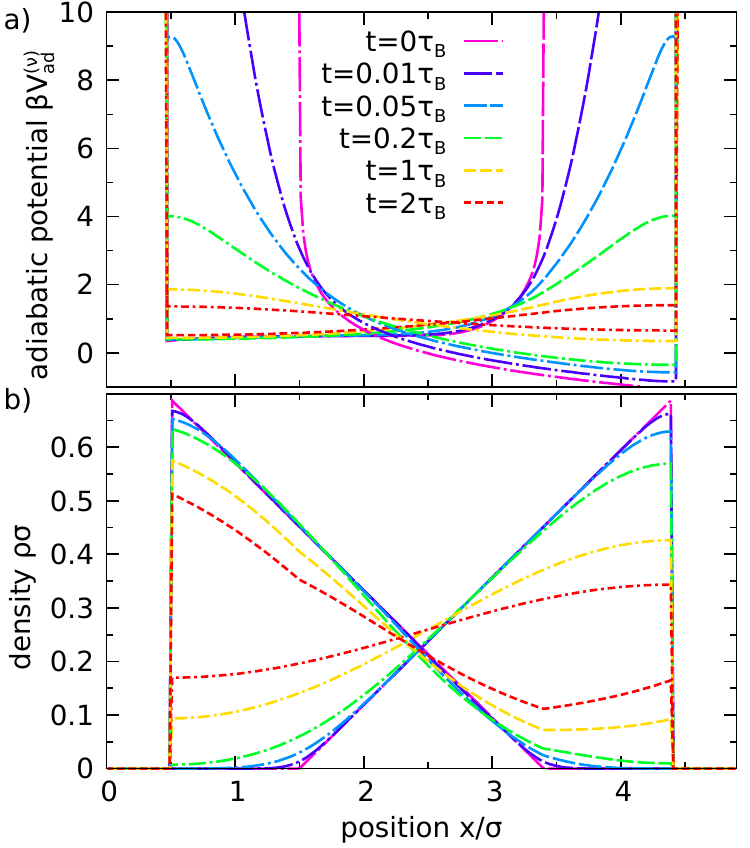}
\par\end{centering}
\protect\caption{(color online)
PCD results for the first component of a species-labeled two-particle system (dashed lines), as described in the caption of Fig.~\ref{fig_N2dyn}(a), compared to a single particle with the same initial state as for the omitted second component (dashed-dotted lines with same dash lengths indicating same times). 
Shown are (a) the adiabatic potentials $V^{(\nu)}_\text{ad}(x,t)$ (shifted vertically for better comparison)
and (b) the density profiles $\rho_{2}^{(1)}(x,t)$ and $\rho_1(x,t)$ (only the first species for $N=2$).
\label{fig_N2dyn2}}
\end{figure}

Now we return to the description of mixtures of identical particles, where the ideal contribution to the free energy is given by Eq.~\eqref{eq_DFrhoID}.
This functional depends explicitly on the individual density profiles and not only on their sum.
Hence, there is only one ``equilibrium'' solution for the densities minimizing the functional.
However, since there is no distinction between the two species in the excess free energy, Eq.~\eqref{eq_DFrhoEX},
the resulting density profiles represent the most disordered state that is compatible with the total interaction.
In the canonical case with $N=2$ particles, this means that the profiles Eq.~\eqref{eq_rho11eqNO}, corresponding to the dashed lines in Fig.~\ref{fig_N2eq},
result in a smaller value of the canonical density functional than in the physical equilibrium state.
As a logical consequence, choosing Eq.~\eqref{eq_rho11eq} as the initial profiles,
the PCD for mixtures spuriously drive the system out of the actual equilibrium, which we show in Fig.~\ref{fig_N2dyn}(a).
This behavior illustrates clearly and already at early times the unphysical artifact observed in Sec.~\ref{sec_results} that the particles tend to mix.
Intriguingly, the PCD for mixtures also spoils the time evolution of the total density profile, which becomes obvious from Fig.~\ref{fig_N2dyn}(b),
where the system, seemingly initiated in equilibrium, exhibits a non-trivial dynamical behavior, just to finally return to a state with the same total density,
but different single-particle profiles.
This is a clear indication that the present form of the PCD do not reproduce the (correct) results of the one-component version~\cite{canonical4}.

The situation described above is somehow reminiscent of the mixing paradox, which
tells us that one should not assign a different entropy to a mixed and a demixed system (separated by a wall) of ideal particles
if one is not able to measure or does not care about the physical difference between two species.
In this case, no entropy change upon mixing or reseparation may occur.
For the mixture of one-dimensional hard rods considered here, 
the particle interactions take the role of a wall inserted into the system.
In inversion of the argumentation for the mixing paradox, we expect a higher entropy (or lower free energy) for the demixed state (true equilibrium), 
where the mixed state should even be entropically forbidden.
This means that, in our theory, we care about a difference that is not reflected by the mathematical 
structure of the PCD for mixtures.
Therefore, the entropy (or the canonical free energy) employed for a mixture is ill-defined.

To resolve this ``inverse mixing paradox'', we continue the discussion from Sec.~\ref{sec_distinguish}.
Both partition functions $Z_{2}$ from Eq.~\eqref{eq_Z2} and $Z_{11}$ from Eq.~\eqref{eq_Z11or} imply a well-defined entropy.
In the DFT language,
the first corresponds to the intrinsic free energy functionals $\mathcal{F}_\text{id}[\rho]$ (all particles distinguishable)
and $\mathcal{F}_\text{ex}[\rho]$ [derived for the symmetric pair potential $u(|x|)$]
of the total density of all particles (species).
The form of $Z_{11}$ [or generally $Z_{N_1N_2}$] should be represented by $\mathcal{F}_\text{id}[\{\rho^{(\nu)}\}]$ (two indistinguishable species)
and an excess term $\mathcal{F}_\text{ex}[\{\rho^{(\nu)}\}]$, explicitly depending on the species-labeled profiles.
Such a (yet unspecified) functional should be based on the asymmetric interaction potential $o(x)\neq o(-x)$ from Eq.~\eqref{eq_o},
which preserves the order of the particles (species), i.e., the non-ergodicity of the system, and thus allows for a physical distinction.
In contrast, the functional introduced in Sec.~\ref{sec_DFT} as the starting point of PCD, 
corresponds to an increased partition function $2Z_2$ [or generally $Z_N N!/(N_1!N_2!)$],
which means that the ideal free energy $\mathcal{F}_\text{id}[\{\rho^{(\nu)}\}]$ implies the combinatorics of two species,
whereas the excess free energy $\mathcal{F}_\text{ex}[\rho]$ is the same as for (orderwise) indistinguishable particles.
More generally, this means that $\mathcal{F}_\text{ex}$ is built from symmetric pair potentials $u(|x|)$ of possibly physically distinguishable particles.
Statically, this overcounting of states does not change the canonical equilibrium density profiles $\rho^{(\nu)}_2(x)$,
ensuring the correct result for the total density $\rho_2(x)$ according to Eq.~\eqref{eq_rho11eqNO}.
In the dynamical case, however, the inconsistency between the entropic force [first term in brackets in Eq.~\eqref{eq_DDFT}, related to $\mathcal{F}_\text{id}$] 
and the interaction force [second term in brackets in Eq.~\eqref{eq_DDFT}, related to $\mathcal{F}_\text{ex}$] 
is the ultimate reason for the wrong time evolution of $\rho_2(x,t)$.

To illustrate the consequences of applying a theory built from symmetric interactions to a non-ergodic system, 
let us consider the time evolution of the adiabatic potential in Fig.~\ref{fig_N2dyn2}(a),
corresponding to the density profiles from Fig.~\ref{fig_N2dyn}(a) of two particles.
Before the first time step, the potential is infinitely steep at the points where the density becomes zero.
This shows that the initial confinement is not intrinsically described by the interaction functional
but has to be artificially generated.
Thus, there is a net force due to $V^{(\nu)}_\text{ad}$ in Eq.~\eqref{eq_DDFTgen}
that drives the dynamics of each particle into the physically forbidden region.
At later times, the generating external fields become less and less restrictive on the interpenetration of the particles 
and become equal (up to a constant) to the external potentials $\Vext^{(\nu)}$ when the equilibrium state of the underlying functional is reached.
It must be the goal to describe the intrinsic interactions in a way that $V^{(\nu)}_\text{ad}(x)\equiv\Vext^{(\nu)}(x)$ in the true equilibrium state.

Finally, we show in Fig.~\ref{fig_N2dyn2}(b) that, at early times and in the low-density regions,
the PCD of one particle in a two-particle mixture is remarkably similar to the proper adiabatic dynamics of a single particle ($N=1$) 
with the initial condition being identical to one of the species-labeled density profiles for $N=2$.
Mathematically, this can be easily explained by the similarity of the adiabatic potentials in the 
regions where the density of the corresponding species is small, compare Fig.~\ref{fig_N2dyn2}(a),
so that also in the two-particle system the main contribution to the free energy stems from the interaction with the generating adiabatic potential. 
This behavior suggests that the speed with which the two profiles mix, which would represent a ``hopping rate'' on the particle level,
is independent of the (local) density.

\section{Discussion \label{sec_discussion}}

In conclusion, we have shown that the presented generalization of PCD to mixtures provides a very good description of the early relaxation dynamics of individual Brownian particles in an interacting system,
in particular, it clearly improves on the grand-canonical DDFT results.
Only at later times, the PCD exhibits an unphysical mixing behavior in one dimension, which dominates 
any deviations arising from the neglect of superadiabatic forces.
Our study of the somewhat artificial one-dimensional case both 
provides valuable insights into more realistic systems in higher spatial dimensions
and is of fundamental theoretical interest in its own right.

We stress that our approach is also relevant for systems much larger than those considered here and even in bulk.
By choosing appropriate external potentials it is possible to isolate a single particle in the initial state.
Therefore, the difference between the canonical and grand-canonical ensemble remains significant for the (single-file) dynamics
even though the differences decrease rapidly for the static properties and joint density of all particles.

To properly describe the non-ergodic Tonks gas, statistical mechanics only constitute a workable approach if an asymmetric interaction potential is considered 
which does not only depend on the relative distance between two particles.
For the present variational approach to work out, one would have to construct a DFT based on such an interaction potential between members of different species.
Such a mixture would then additionally be non-additive,
a case in which even for a symmetric potential in one dimension only an approximate hard-rod functional can be derived~\cite{schmidtNAM}.

Having an accurate ergodicity-breaking theory in one dimension would be very instructive, since one could gain more explicit insight into superadiabatic contributions on the single-particle level. 
A more general and formally exact variational approach, based on a mixture of DFT methods and statistics intrinsically respecting the particle order, 
will be presented in a future publication.
Another promising route would be to analyze the exact results for hard-rod dynamics 
provided by Lizana and Ambj\"ornsson~\cite{LA1,LA2} to see if their lengthy analytical expressions obtained by a Bethe ansatz
can be reformulated within the context of a variational approach.
It will also be insightful to provide a simulation setup that can reproduce the mixed 
canonical density profiles in equilibrium, which will be a challenging task for both BD and (dynamic) Monte Carlo.
This would allow to 
compare the adiabatic-superadiabatic splitting that applies to the artificial case prescribed by PCD
with the (known~\cite{FortiniPRL}) splitting that applies to the true BD case with preserved particle order.

In higher dimensions, the problem described above is seemingly resolved.
Most importantly, these systems are, in general, ergodic, so that the final equilibrium state of PCD is the same as in BD
and the pair potential must also be symmetric.
We showed, however, that this does not guarantee a correct description of the transient dynamical regime [cf.\ Fig.~\ref{fig_N2dyn}(b)]. By contrast it is very likely that there are still some artifacts in the PCD due to its insensitivity to dynamical caging effects.
For example, one would expect in BD that the mixing of the particle-resolved density profiles becomes increasingly slow with increasing total density,
which is not reflected by the conclusions drawn in one dimension from Fig.~\ref{fig_N2dyn2}(b).
More specifically we hold these artifacts accountant for the fact, that density-based theories in general overestimate long-time diffusion constants~\cite{stopper_long}, as it adds unphysical particle exchange dynamics to the physical circuiting of particles, which is slower in dense suspensions of any dimension.
In this sense, we have performed a minimalistic but extreme test (with infinite circuiting time) for the caging scenario in higher dimensions.
It will be challenging to study in detail how significantly this shortcoming of PCD would influence the adiabatic dynamics, since there exists no exact grand-canonical functional for interacting systems in higher dimensions.
Moreover, to describe caging statistically, one would require a complex many-body interaction (and not only an asymmetric pair potential),
which leaves not much hope for a theoretical implementation.
Despite these caveats, the available approximate forms of DFT in three dimension are very accurate~\cite{FMT,Tar99,HaRo06}, so that
we expect that the corresponding PCD will provide a pretty good account for the early (adiabatic) dynamics of single-particle profiles, especially at a low overall density.

Exceptions to the above are presented by glassy, jammed or otherwise arrested systems. 
In such cases the interparticle coupling is so strong (due to, e.g., high density or strong attractions) that the phase space can not be fully explored and canonical averaging is not 
appropriate. In particular, for high-density systems with purely repulsive interactions the 
cage effect is a dominant physical mechanism.
The failure of the `three-rod-caging test' in Fig.~\ref{fig_rho_of_tb} ultimately points to the impossibility 
of describing such glassy states using theories for which the density is the only variable. 
The standard observable used to quantify the dynamics of arrested states is the van Hove 
function, the self part of which describes the dynamics of a single tagged particle. 
DDFT has been employed in the test particle limit to approximate this self van Hove 
function~\cite{dtp,stopper_bulk}. In the light of our present study one can safely conclude 
that the dynamic arrest which has been observed in DDFT calculations is an artifact arising 
from an approximate free energy functional~\cite{dtp}, rather than a true indication of vitrification.

To provide a more sophisticated description of caging effects on the level of symmetric pair potentials in any dimension, it seems unavoidable that 
one must extend the variational approach beyond the one-point density alone.
For example, variational approaches based on two-point correlations might be better able to cope 
with caging.
Alternatively, to obtain improved results using the one-point density alone in a non-variational framework, it may be possible 
to incorporate superadiabatic effects by relaxing the requirement 
of time locality, i.e., incorporate memory functions. 
By far the most natural extension of DDFT (in the authors opinion) lies in the framework of power 
functional theory~\cite{pft1,pft2}, based on a functional of both density {\it and} current, which is 
nonlocal in both space and time. 
Indeed, it seems clear that a vector field is necessary in order to describe the motion of a 
fluid. 
So far, workable approximations to the power functional have remained time-local~\cite{pft2}. 
It remains, however, unclear, whether implementing superadiabatic effects will automatically provide a better description of caging,
which, as laid out in this paper, can be understood as an adiabatic many-body effect,
for the description of which we require an accurate treatment of the static interactions.

\begin{acknowledgments}
	T.S. was supported by the Deutsche Forschungsgemeinschaft as part of the Forschergruppe GPSRS under grant ME1361/13-2.
\end{acknowledgments}

\appendix
\section{Brownian dynamics simulations \label{app_BD}}

For calculation of trajectories in our BD simulations we employ a hybrid algorithm. It treats random forces and soft external potentials in discrete time steps according to the standard BD approach. The interactions between particles or between particles and walls, however, are treated as instantaneous collisions to account for the hard core repulsion.
We chose this approach over the usual approximation of hard core interactions with steep soft potentials to avoid smoothing effects and instead obtain exact hard core density profiles.

In each time step of length $\dt=5\times10^{-5}\tau_\mathrm{B}$ we calculate the instantaneous velocity $v_i(t)$ of a particle $i=1,...,N$ (of species $\nu$),
which, in overdamped Langevin dynamics, directly adjusts to the force $D_0\beta \bvec{f}_i(t)= v_i(t)$ acting on the particle.
The force reads
\begin{equation}
\bvec{f}_i(t)=-\frac{\partial}{\partial x_i}V_{\text{ext},0}^{(\nu)}(x_i)\Theta(-t) + R_i(t),
\end{equation}
where $R_i(t)$ is a random force drawn from Gaussian white noise with variance $2/(D\beta^2 \dt)$ and zero mean, the particular harmonic trap potentials $V_{\text{ext},0}^{(\nu)}$ 
are specified in the caption of Figs.~\ref{fig_rho_of_t} and~\ref{fig_rho_of_tb} and the Heaviside step function represents the switch-off of the harmonic trap at $t=0$.
From the old position $x_i(t)$ and the velocity we can compute the preliminary new position via
\begin{equation}
\tilde{x}_i(t+\dt)=x_i(t) + v_i(t) \dt.
\end{equation}
If particles overlap with each other or the walls, then we resolve these overlaps in billiard-like collisions. For each created overlap we first calculate the collision time,
\begin{equation}
t_i=\frac{x_{i+1}(t)-x_i(t)-\sigma}{v_i(t)-v_{i+1}(t)},
\end{equation}
for a collision of particles $i$ and $i+1$, or
\begin{equation}
t_i=\frac{x_\mathrm{w}-x_i(t)}{v_i(t)},
\end{equation}
for a particle-wall collision, where $x_\mathrm{w}=R,L-R$ is the minimum or maximum position of a particle in the slit.
Then, these collisions are scheduled and resolved chronologically to obtain the final positions of the time step via
\begin{align}
x_i(t+\dt)=\,&\tilde{x}_{i+1}(t+\dt) - \sigma,\\
x_{i+1}(t+\dt)=\,&\tilde{x}_{i}(t+\dt) + \sigma
\end{align}
for particle particle collisions or
\begin{equation}
x_i(t+\dt)=2x_\mathrm{w}-\tilde{x}_i(t+\dt)
\end{equation}
for particle-wall collisions.
(If hereby new overlaps are created, then the collisions, again, have to be scheduled and resolved.)
When all collisions are resolved, the algorithm resumes with calculating the velocities for the next time step.

To sample the time dependent density profiles we generate an equilibrated configuration with harmonic traps switched on and an initial equilibration time of $1\tau_\mathrm{B}$.
Between each sample we equilibrate the initial configuration for another $10^{-3}\tau_\mathrm{B}$ to decorrelate the samples from each other.
A sample is then taken by switching off the harmonic traps and relaxing the system to its final state.
The density profiles of the species shown in Sec.~\ref{sec_results} are obtained by averaging over $4\times10^6$ individual samples via
\begin{equation}
\rho_{N_1 N_2}^{(\nu)}(x,t)=\left\langle\sum_{i}^{N_\nu}\delta(x_i(t)-x)\right\rangle,
\end{equation}
where the sum runs over all particles belonging to the respective species.

\end{document}